# Magnetic and Dielectric Properties of Multiferroic $BiFeO_3$ Nanoparticles Synthesized by a Novel Citrate Combustion Method

**Samar Layek\* and H. C. Verma**

Department of Physics, Indian Institute of Technology, Kanpur 208016, India.

**Abstract**

Single phase $BiFeO_3$ nanoparticles have been successfully synthesized for the first time by a novel citrate combustion method without using any solvent. Well mixed metal nitrates along with citric acid which is used as fuel combust to give $BiFeO_3$ nanoparticles after annealing. These particles are single phase in nature and crystallize in the rhombohedral distorted perovskite structure (space group-*R3c*) which has been confirmed by the Rietveld refinement of the room temperature powder x-ray diffraction data. Nearly spherical particles of average particle size 47 nm have been seen from transmission electron micrograph. Room temperature magnetic hysteresis measurement shows weak ferromagnetism though the magnetization does not saturate upto 1.75 T applied field. The coercive field value is calculated to be 180 Oe which is 3 times higher than that prepared by solvent free combustion method using Glycine. $^{57}Fe$ Mössbauer spectrum can be fitted with a sextet corresponding to single magnetic state of hyperfine field about 49.5 T corresponding to $Fe^{3+}$ state of the iron atom. The dielectric relaxation and ac conductivity as a function of frequency have been discussed. High dielectric permittivity has not been found in these nanoparticles like other reported $BiFeO_3$ ceramics.

## 1. INTRODUCTION

Multiferroic materials are compounds which show more than one ferroic or anti-ferroic properties (like ferromagnetic, ferroelectric, ferroelastic etc,) in the single phase sample. In these material ferroic properties are coupled with each other [1]. Among these multiferroics, the materials possessing ferroelectric (anti-ferroelectric) and ferromagnetic (anti-ferromagnetic) properties are called magnetoelectric compound. In a magnetoelectric compound, application of external electric field can induce magnetization and application of magnetic field can induce intrinsic polarization. The coupling between these properties for a magnetoelectric compound makes them important not only for industrial application but also from physics point of view because of their enriched physical properties [2, 3]. But unfortunately there are very few materials existing in the nature or synthesized in the laboratory which show both these ferroic properties at or above room temperature.

Bismuth ferrite ($BiFeO_3$) is one of the examples of the multiferroic material which crystallizes in the distorted perovskite structure. It shows G-type antiferromagnetic ordering below Neel temperature $T_N$ = 370 $^0C$ and ferroeletricity below $T_C$ = 830 $^0C$ [4, 5]. In addition to this antiferromagnetic ordering, it shows modulated spin spiral structure of periodicity 62 nm. $BiFeO_3$ in the bulk form shows very low magnetization due to this spin spiral structure [6]. Formation of impurity phases during preparation is another drawback with $BiFeO_3$ based systems. From Bi-Fe-O phase diagram it is known that the 113 perovskite phase forms near 800 $^0C$ which is well above the melting point of Bi (≈271 $^0C$) [7]. Iron rich impurity phases, like $Bi_2Be_4O_9$ etc are often found to evolve along with pure perovskite phase due to Bi loss during preparation of $BiFeO_3$ bulk material. These drawbacks of this material are preventing for its real use in industrial application. These difficulties can be fully or partially removed by doping on both Bi and Fe site or by preparing nanosized materials at low temperature. Enhancement of the magnetic properties in the single phase $BiFeO_3$ ceramics has been found by A-site (Bi site) doping of both 3+ rare earth atoms such as (La, Nd, Y etc) and 2+ alkaline earth atoms (Ca, Ba, Sr etc) [8-13]. B site (Fe-site) doping of transition metal elements like Cr, Mn, Ti, Zr etc also resulted into higher magnetization value [14-17].

The structural and magnetic properties of $BiFeO_3$ nanoparticles greatly depend on the particle size, morphology, doping, pressure etc, which can be controlled by the synthesis procedure. So, preparation method has pronounced effect on the structural and magnetic properties of $BiFeO_3$ nanoparticles. $BiFeO_3$ nanoparticles have been synthesized by different techniques such as 1) glycine nitrate combustion method [18], 2) sol-gel method [19], 3) tartaric acid [20] 4) sucrose [21] based method etc. In many of these methods water is being used as a solvent for preparing the solution from metal salts, mostly nitrates. But, bismuth sub-nitrate (Bi(NO$_3$)$_3$, 6H$_2$O) is not soluble in water and turns into oxy-nitrate which may result into impurity phases. A small amount of Fe$_3$O$_4$ phase has been found along with pure BiFeO$_3$ nanoparticles by glycine nitrate combustion method [18]. Nanoparticles of BiFeO$_3$ have been seen not only to improve the magnetization but also useful in many applications like visible light photocatalytic activity [22], microwave absorption [23], gas sensing properties [24], etc.

BiFeO$_3$ also shows interesting dielectric properties. Recently Hunpratub et al. [25] have found high dielectric permittivity ($\approx 10^3$-$10^4$) in the wide range of temperature -50 to 200 $^0$C in BiFeO$_3$ ceramic prepared by precipitation method. Three kinds of dielectric relaxations have been found. Debye-type relaxation arising from the carrier hopping process between Fe$^{2+}$ and Fe$^{3+}$ atoms have been seen in the temperature range -50 to 20 $^0$C. At high temperatures, other two kinds of dielectric relaxations have been found due to grain boundary effect and defect ordering. Non-linear dielectric properties with high dielectric permittivity value ($\approx 10^4$) above room temperature have been found in BiFeO3 ceramic prepared by solid state reaction method [26].

In this paper, we present our work on synthesis of pure phase BiFeO$_3$ nanoparticles prepared by a novel combustion method using metal nitrates and citric acid in absence of solvent. The crystal structure and morphology of the prepared nanoparticles are studied using x-ray diffraction (XRD) and transmission electron microscopy (TEM). Magnetic characterizations are done using vibrating sample magnetometer (VSM) and $^{57}$Fe Mössbauer spectroscopy. The frequency variation of dielectric properties is also studied in order to understand the electrical properties.

## 2. EXPERIMENTAL DETAILS

### 2.1. Material Synthesis

BiFeO$_3$ nanoparticles have been prepared by a combustion method using metal nitrates and citric acid but without using any solvent. Bismuth nitrate (Bi(NO$_3$)$_3$, 5H$_2$O), ferric nitrate (Fe(NO$_3$)$_3$, 9H$_2$O) and citric acid (C$_6$H$_8$O$_7$) all were used with purity 99.9% or higher and without further purification. These compounds were taken in stoichiometric ratios in a glass beaker and heated at 80 $^0$C on a hot plate for about 20 minutes with constant stirring with cleaned glass rod to mix them together. Then the mixture was heated at 150 $^0$C for combustion of the mixture to form brown colour precursor. The precursor was grounded into powder using mortar and pastel. Precursor powder was annealed using high purity alumina crucibles in air at 600 $^0$C for 2 hours inside a programmable box furnace with heating rate 100 $^0$C per hour and then slowly cooled down to room temperature. The chemical reaction for this synthesis is given in equation 1.

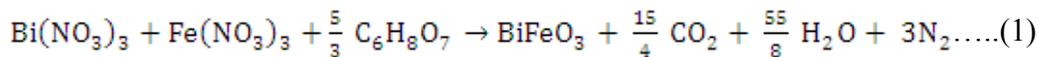

$$Bi(NO_3)_3 + Fe(NO_3)_3 + \tfrac{5}{3} C_6H_8O_7 \rightarrow BiFeO_3 + \tfrac{15}{4} CO_2 + \tfrac{55}{8} H_2O + 3N_2 \ldots(1)$$

Bulk BiFeO$_3$ sample was prepared by standard solid state reaction method using high purity oxides in order to compare the magnetic properties. We have also prepared the single phase BiFeO$_3$ nanoparticles by solvent free combustion method using glycine (NH$_2$CH$_2$COOH) which has also been used as a fuel for the combustion [27]. The heat of combustion and decomposition temperature of glycine (13.0 kJ/gm and 262 $^0$C) are higher than that of citric acid (10.2 kJ/gm and 175 $^0$C) [28]. The single phase sample prepared by citric acid combustion needs lower temperature (150 $^0$C) than those using glycine as fuel (180 $^0$C) [27].

**2.2. Characterization Tools**

Small amount of annealed powder was characterized by XRD on ARL X'TRA X-ray diffractometer (Thermo Electron Corporation) using Cu-K$\alpha$ radiation in order to check the phase purity and calculate average crystallite size. Crystallite size was calculated from the Scherrer formula using the peak broadening of XRD peaks after correcting the instrumental broadening. Small amount of powder was dispersed in alcohol (99.99% pure) and was sonicated for 15 minutes. A small drop of this solution was dropped on a Carbon coated copper grid and was used for TEM characterization using a FEI Tecnai G2 electron microscope operated at 200 kV. Room temperature isothermal magnetization data was taken using a vibrating sample magnetometer (VSM, ADE Technologies, USA) upto applied field of 1.75 T. Room temperature Mössbauer spectrum was recorded by using a standard constant acceleration $^{57}$Fe Mössbauer spectrometer in the transmission geometry in which a 25 mCi $^{57}$Co in rhodium matrix was used as the source. Collected data were analyzed by least square fit program considering Lorentzian line-shape of the spectrum. Pure iron spectrum was used for calibration. The nanoparticle sample was pressed into a pellet of 8 mm diameter and about 1 mm thickness and sintered at 600 $^0$C for 2 hours in a box furnace with controlled heating. Silver paste was put on the flat surface of the sintered pellet for electrical measurements. Electrical measurements as a function of frequency were performed by using Agilent E4980A precision LCR meter in the frequency range 20Hz-2MHz at room temperature.

## 3. RESULTS AND DICUSSIONS

### 3.1. Structural Analysis

Room temperature powder x-ray diffraction pattern of the BiFeO$_3$ nanoparticles along with Rietveld refinement is presented in Figure 1. All the peaks could be indexed in the rhombohedral distorted perovskite structure with space group-*R3c*. No evidence of impurity phases (like Fe$_2$O$_3$, Fe$_3$O$_4$, Fe, Bi$_2$Fe$_4$O$_9$ etc,) has been found at least to the detection limit of XRD. The average crystallite size calculated from the XRD peak broadening after correction from the instrumental broadening is about 47±5 nm. XRD diffraction data in 2θ range 20$^0$-60$^0$ has been used for the refinement. Bi atoms at x=0, y=0 and z=0, Fe atoms at x=0, y=0 and z=0.2194 and oxygen atoms at x=0.4346, y=0.0119 and z=-0.0468 crystallographic positions in the rhombohedral structure (space group *R3c*) were taken as the starting parameters for the refinement. The lattice parameters a=b=5.571 Å and c=13.858Å (taken from single crystal data from the reference [29]) were taken as the initial values and then refined for the intensity matching. The crystal parameters are calculated to be a= 5.5725 Å and c= 13.8437 Å from the refinement of the XRD data.

Figure 2 shows the TEM micrograph for BiFeO$_3$ nanoparticles synthesized by citric acid based combustion method. A broad particle size distribution with the diameter of the particles ranging from 35 to 65 nm can be seen by detailed study. These particles are nearly spherical in shape.

The particles are not well separated from each other and agglomerations of particles have been observed at places.

**3.2. Magnetic Properties**

Figure 3 shows the field variation of the magnetization (hysteresis curve) for $BiFeO_3$ nanoparticles prepared by citric acid combustion method along with its bulk counterpart prepared by solid state reaction. The highest applied magnetic field is 1.75 T. Pure $BiFeO_3$ in the bulk form is an antiferromagnet below 370 $^0$C. The decrease in the particle size leads to increase in the magnetization. The magnetic loop for bulk sample shows very low magnetization with nearly zero hysteresis. The hysteresis curve for 47 nm nanoparticles show a typical weak ferromagnetic like behavior where the magnetization does not attain its saturation value upto the highest applied magnetic field. The coercive field value calculated from the hysteresis loop is about 180 Oe which can also be seen from the inset of figure 3. The value of magnetization in these nanoparticles is about double of that in bulk size sample.

The increase in the magnetization as the size goes down may arise because of partial suppression of the spin spiral ordering due to finite size effect of the prepared nanoparticles. The value of magnetization in our case is lower than that of the typical $BiFeO_3$ nanoparticles reported in [18, 20]. Water was used as solvent during preparation of those nanoparticles which could result into higher magnetization coming from impurity phases as discussed in [18]. We have not used any solvent during the preparation of these nanoparticles so there is less probability of getting impurity phases (like $Fe_2O_3$, $Fe_3O_4$, Fe, $Bi_2Fe_4O_9$ etc,). The absence of impurity phases has been confirmed by slow scan XRD measurement. Though, there is not much difference in the magnetization value of these nanoparticles with that prepared by glycine-nitrate combustion method, coercive field value is found to be 3 times higher [27]. The increase in the coercive field value is a signature of increase in the magneto-crystalline anisotropy which may arise due to broad particle size distribution which can be seen from TEM image.

**3.3. Mössbauer Spectrum Study**

$^{57}$Fe Mössbauer spectroscopy is one of the most novel and efficient tools to investigate the local magnetic behavior and oxidation state of the iron atoms in the matrix. The Mössbauer spectrum for these nanoparticles is shown in figure 4. The black dots represent the experimentally recorded data points whereas the blue lines are the least square fit of the spectrum. The experimental data can be fitted with a sextet with isomer shift of about 0.36±0.02 mm/s and hyperfine field of 49.5±0.1 T corresponding to a single magnetic state of Fe-atoms. A six line magnetic sextet at room temperature implies that the magnetic ordering temperature is well above room temperature. No impurity phases (like $Fe_2O_3$, $Fe_3O_4$, Fe, $Bi_2Fe_4O_9$ etc,) have been found which support our XRD result. $^{57}$Fe Mössbauer spectroscopy is an efficient tool for detecting small iron containing phases. The absence of the other iron containing phases implies that the increase in magnetization is intrinsic in nature coming from *R3c* phase of the $BiFeO_3$ nanoparticles. The isomer shift of 0.36 mm/s indicates that Fe atoms are in 3+ states in the $BiFeO_3$ matrix. The hyperfine field is calculated to be 49.5 T for these strongly interacting particles which is similar to the previous reported values [30]. Single magnetic sextet has also been found in Mn-doped $BiFeO_3$ ceramics [31]. There are some reports where the spectrum is fitted with two magnetic sextets with nearly same hyperfine field value in some $BiFeO_3$–based compounds [11, 32]

### 3.4. Dielectric Properties

Dielectric constant (ε') and loss tangent (Tan(δ)) of the sintered pellet of the $BiFeO_3$ sample prepared by combustion method have been measured as a function of frequency at room temperature. The variations of both dielectric constant and loss tangent are shown in figure 5 as a function of frequency in the range 20 Hz- 2MHz. Both these quantities decrease with increasing frequency. The low frequency region indicated the presence of dc conductivity [33, 34]. In high frequency range both dielectric permittivity and loss factor decrease slowly and nearly attain constant values. At 1 MHz, the value of dielectric constant is ≈ 57±2 and loss factor is less than 2%. Similar kind of dielectric behavior has been found in undoped and doped $BiFeO_3$ ceramics. The value of dielectric constant, in this particular frequency range, of our sample is comparable or better than some of the $BiFeO_3$ based systems [10, 12, 14, 35]. But the value of dielectric permittivity are small compared to BiFeO$_3$ ceramics (≈10$^3$-10$^4$) prepared by precipitation and solid state reaction methods [25, 26]. This may be due to the large amount of porosities in our nanoparticles samples which is annealed at 600 $^0$C. BiFeO$_3$ ceramics annealed at 600 $^0$C exhibits low value of ε' (≈ 100) whereas 700 $^0$C annealed ceramics exhibits very high value of ε' (≈ 2000) at low frequency as later shows very dense microstructure with distinct grain and grain boundary structure [25]. So, the grain and grain boundary structure has significant impact on the electrical properties for such systems.

Figure 6 shows the frequency variation of the ac conductivity of the nanoparticles measured at room temperature. At low frequency range the ac conductivity is nearly independent of the frequency and has been attributed to the dc conductivity of the sample. With increasing frequency the ac conductivity increases, at first slowly and then rapidly at about 100 KHz. The ac conductivity can be ascribed by the relation given in equation (2)

$$\sigma_{ac} = \sigma_{dc} + Af^{\alpha} \qquad \ldots\ldots\ldots (2)$$

where $\sigma_{dc}$ is the dc conductivity of the material, A and α are constants. The fitted curve according to the equation (2) is shown by solid blue lines in figure 6. The value of the dc conductivity as obtained from the fit are $\sigma_{dc}$ = 2.68 x 10$^{-7}$ S/m whereas the value of the exponent, α =0.741. Similar kind of value of the dc conductivity has been found in alkaline doped BiFeO$_3$ ceramics at room temperature [36].

## 4. CONCLUSIONS

BiFeO$_3$ nanoparticles with average crystallite size of about 50 nm have been prepared by a simple combustion method using metal nitrates and citric acid. X-ray diffraction data shows that nanoparticles are single phase in nature and crystallize in the same structure (distorted perovskite of space group *R3c*) as the bulk compound. Magnetization with about twice of value of bulk sample has been found in VSM measurement. The large increase in the magnetization may be resulting from the suppression of the spin spiral structure due to nano size. Room temperature Mössbauer spectrum also confirmed that the increase in magnetization is intrinsic and not resulting from any impurity phases. Frequency variation of the dielectric and ac conductivity has been investigated.

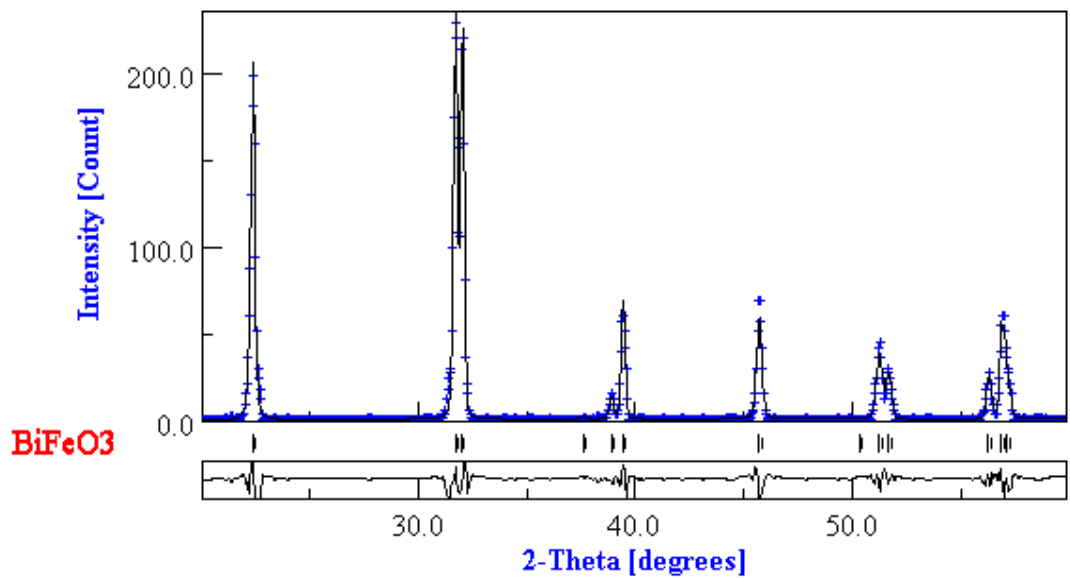

**Figure 1:** Rietveld refinement profile of the powder XRD pattern of BiFeO$_3$ nanoparticles at room temperature. Blue dots are the experimental points, solid black is line fit, in-between black bars are Bragg position of the BiFeO$_3$ crystallizes in space group-*R3c* and difference between experimental data and theoretical fit are shown below.

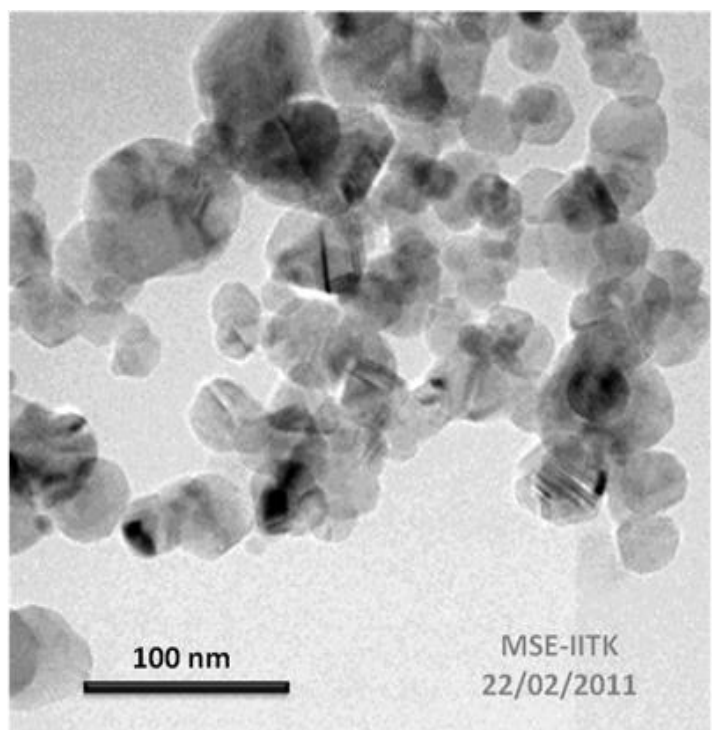

**Figure 2:** TEM micrograph BiFeO$_3$ nanoparticles prepared by combustion method using citric acid.

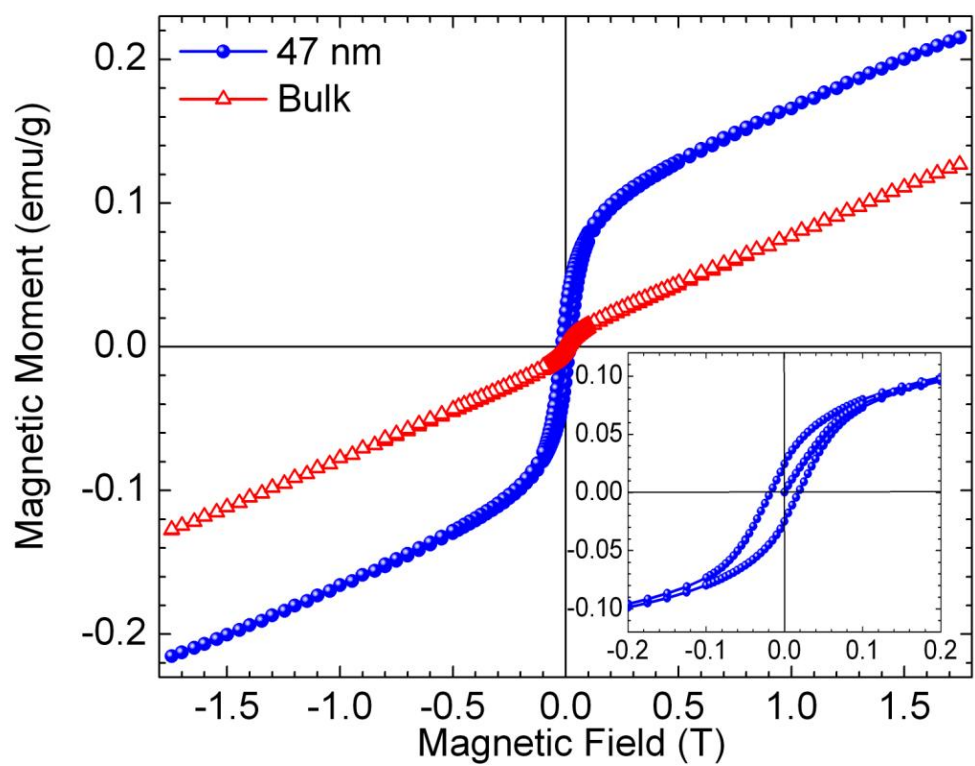

**Figure 3:** Magnetization as function of applied field (hysteresis loop) has been plotted for the BiFeO$_3$ nanoparticles measured at 300 K by VSM. (Inset) same hysteresis loop in the low field region.

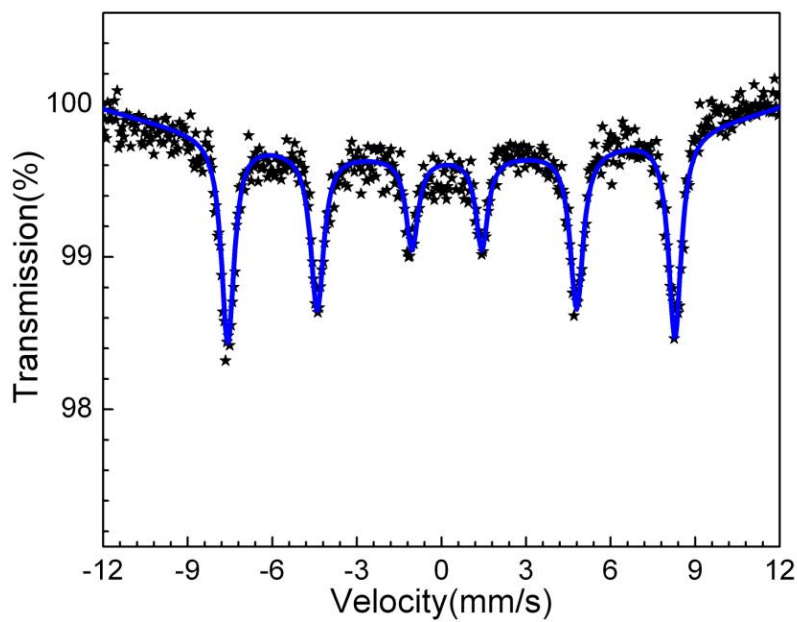

**Figure 4:** Mössbauer spectrum of BiFeO$_3$ nanoparticles prepared by combustion method measured at 300K.

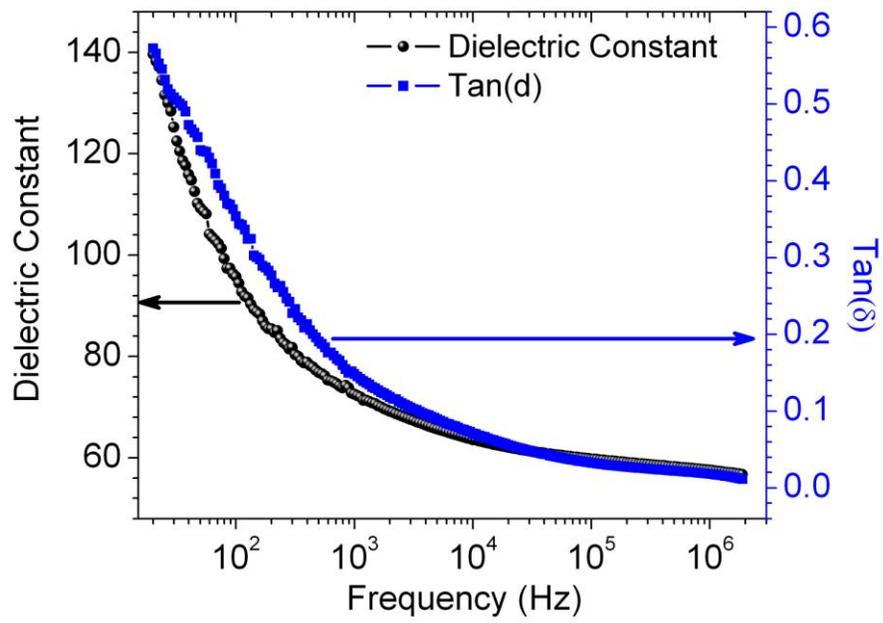

**Figure 5:** Dielectric constant (ε′) and loss tangent (Tan(δ)) as a function of frequency of $BiFeO_3$ nanoparticles prepared by combustion method measured at 300K.

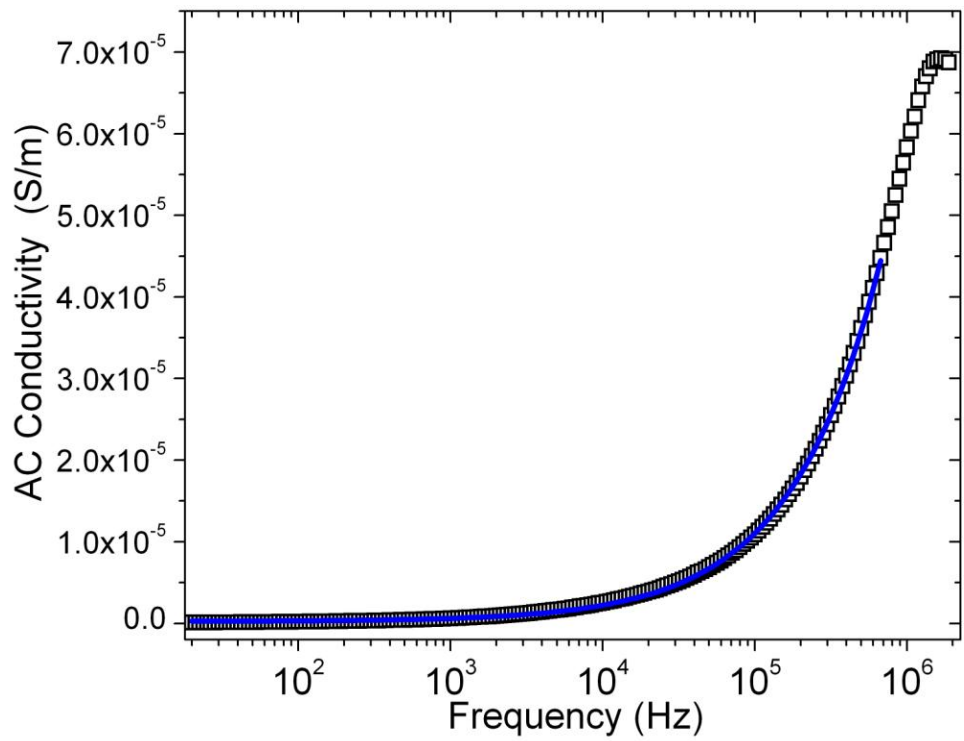

**Figure 6:** Room temperature ac conductivity ($\sigma_{ac}$) as a function of frequency of $BiFeO_3$ is plotted and blue solid line is the fit according to equation (2).